\PassOptionsToPackage{table}{xcolor}
\documentclass[sigconf,screen]{acmart}
\nonumber
\usepackage{booktabs} 
\usepackage{graphicx} 
\usepackage{amsmath}  
\usepackage{url}      
\usepackage{subcaption} 
\definecolor{gold}{RGB}{255,215,0}  

\usepackage[utf8]{inputenc}  
\usepackage[greek,english]{babel}  

\AtBeginDocument{%
  }

\acmConference[Working paper]{Make sure to enter the correct
  conference title from your rights confirmation emai}{Jan.
  2024}{Metaverse}

\setcopyright{none}
\renewcommand\footnotetextcopyrightpermission[1]{}
\begin{document}

\title{EthosGPT: Mapping Human Value Diversity to Advance Sustainable Development Goals (SDGs)}
\author[Luyao Zhang*]{Luyao Zhang}
\affiliation{%
  \department{Social Science Division and Digital Innovation Research Center}
  \institution{Duke Kunshan University}
  \country{China}
}

\authornote{Corresponds to: Luyao Zhang (email: lz183@duke.edu, address: Duke Kunshan University, No.8 Duke Ave. Kunshan, Jiangsu 215316, China.) }


\begin{abstract}
Large language models (LLMs) are transforming global decision-making and societal systems by processing diverse data at unprecedented scales. However, their potential to homogenize human values poses critical risks, akin to biodiversity loss undermining ecological resilience. Rooted in the ancient Greek concept of \textit{ēthos}—denoting both individual character and the shared moral fabric of communities—\textbf{EthosGPT} draws on a tradition that spans from Aristotle’s virtue ethics to Adam Smith’s moral sentiments as the ethical foundation of economic cooperation. These traditions underscore the vital role of value diversity in fostering social trust, institutional legitimacy, and long-term prosperity. \textbf{EthosGPT} addresses the challenge of value homogenization by introducing an open-source framework for mapping and evaluating LLMs within a global scale of human values. Leveraging international survey data on cultural indices, prompt-based assessments, and comparative statistical analyses, EthosGPT reveals both the adaptability and biases of LLMs across regions and cultures. It offers actionable insights for developing inclusive LLMs, such as diversifying training data and preserving endangered cultural heritage to ensure representation in AI systems. These contributions align with the United Nations Sustainable Development Goals (SDGs), especially \textbf{SDG 10 (Reduced Inequalities)}, \textbf{SDG 11.4 (Cultural Heritage Preservation)}, and \textbf{SDG 16 (Peace, Justice and Strong Institutions)}. Through interdisciplinary collaboration, EthosGPT promotes AI systems that are both technically robust and ethically inclusive—advancing value plurality as a cornerstone for sustainable and equitable futures.
\end{abstract}




\begin{teaserfigure}
  \includegraphics[width=\textwidth]{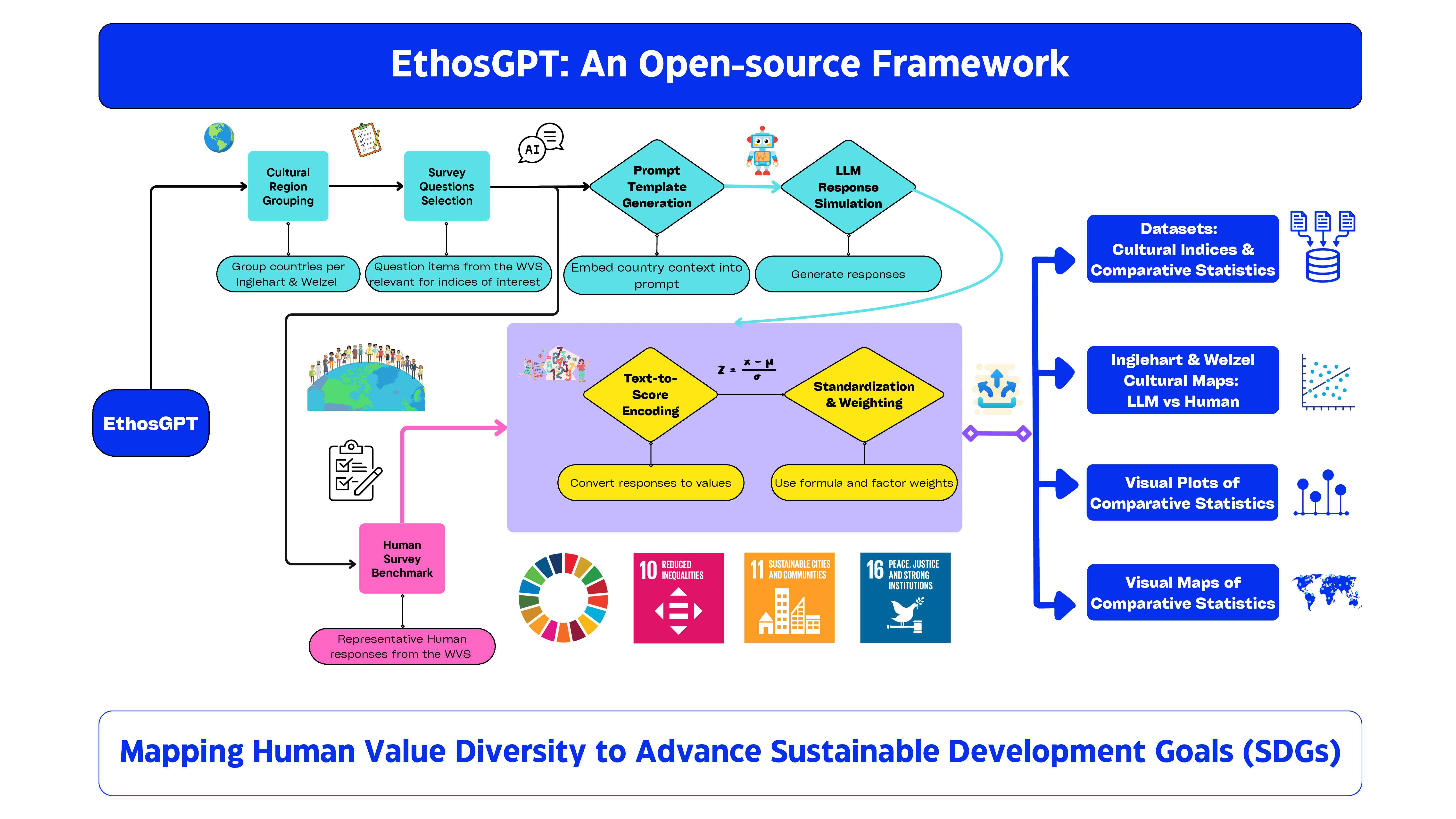}
  \caption{EthosGPT: An Open-Source Framework for Mapping Human Values Across Cultures in Large Language Models.}
  \label{fig:teaser}
\end{teaserfigure}


\maketitle

\section{Introduction}

Large language models (LLMs) are transforming global decision-making and societal systems by processing diverse data at unprecedented scales~\cite{zhao2024surveylargelanguagemodels}. However, their increasing influence on culture, communication, and policy raises urgent questions about how such models reflect — or overwrite — the plurality of human values~\cite{xu2023cvalues, li2024culturellm}. As LLMs gain the power to simulate agents, shape discourse, and inform governance, the risk of homogenizing values becomes critical — a sociotechnical equivalent of biodiversity loss undermining ecological resilience~\cite{10.1093/rof/rfae010,10.1093/nsr/nwab032,Pascual2021,Diaz2022}. Societies, much like ecosystems, thrive through diversity — of perspectives, moral systems, and cultural expressions — which are essential for adaptability, innovation, and long-term prosperity.

To address this challenge, we introduce \textbf{EthosGPT} — an open-source framework for mapping and evaluating LLMs within a global landscape of human values. The name \textit{EthosGPT} draws inspiration from the ancient Greek term \textgreek{ἦθος}, which in classical philosophy denotes not only individual character or disposition, but the shared moral nature and customs that bind communities~\cite{aristotle2018rhetoric}. Aristotle emphasized \textit{ēthos} as a mode of persuasion grounded in virtue and credibility~\cite{aristotle2018rhetoric}— a deep form of ethical resonance rooted in lived experience and communal identity. 

This classical understanding finds an echo in the foundational principles of modern economics. Adam Smith, often regarded as ``the father of economics'', began not with markets but with morality. In \textit{The Theory of Moral Sentiments}~\cite{Haakonssen2002-HAAAST-2}, Smith argued that human behavior is governed by an innate capacity for empathy — what he called ``fellow-feeling'' — which enables individuals to imagine themselves in the situation of others. This affective and ethical capacity, he believed, was the necessary substrate for justice, trust, and ultimately, economic cooperation. Thus, before \textit{The Wealth of Nations}~\cite{smith2014wealth} could envision efficient markets, it rested on the premise of ethical intersubjectivity: the ability to feel with others and recognize moral plurality.  Modern literature on corporate culture similarly emphasizes the normative foundations of cooperation and performance in economic settings~\cite{Guiso2022Culture}.
The purpose of EthosGPT is to operationalize the philosophical and ethical commitment of ethos in a computational setting.

In this spirit, EthosGPT aims to interrogate and shape how LLMs reflect and negotiate human values across cultural boundaries — not simply through logical coherence (\textit{logos}) or emotional resonance (\textit{pathos}), but through ethical depth and contextual relevance (\textit{ēthos}), as the three persuasive appeals were first articulated in Aristotle’s Rhetoric~\cite{aristotle2018rhetoric}. Specifically, it explores how LLMs can simulate culturally grounded agents and how their responses align or diverge from human cultural data. This framework is guided by two central research questions:

\begin{itemize}
    \item \textbf{RQ1:} How can we design a general, open-access framework that leverages LLMs to simulate representative cultural agents for cultural entities across diverse cultural indices?
    \item \textbf{RQ2:} In what ways do cultural indices derived from LLM-simulated national agents differ from those based on human responses in global survey data across culturally diverse regions?
\end{itemize}

To address these questions, EthosGPT employs a dual-methodology approach:
\begin{enumerate}
    \item \textbf{Mapping Cultural Indices through Prompt-Based Assessments with Measures and Visualizations:} Drawing on data sources such as the World Values Survey~\cite{haerpfer2022wvs, inglehart2005modernization}, EthosGPT constructs prompt-based assessments to elicit LLM-generated responses representing different cultural profiles. These responses are evaluated and visualized to examine how effectively LLMs serve as representative agents, addressing \textbf{RQ1}.
    \item \textbf{Comparative Statistical Analyses and Visualizations:} Addressing \textbf{RQ2}, we employ statistical techniques and visual tools to compare LLM-generated cultural indices with those derived from empirical human survey data~\cite{tao2024cultural}, identifying alignment, discrepancies, and potential biases.
\end{enumerate}

This dual-method approach ensures a comprehensive framework linking qualitative cultural representation with quantitative validation. The results of this methodology are twofold:
\begin{itemize}
    \item \textbf{R1:} An open-source framework of how LLMs can act as cultural agents by aligning their output with known survey-derived indices.
    \item \textbf{R2:} A systematic analysis of where and why LLM-generated cultural indices diverge from human populations, revealing biases, gaps, or limitations in model training and design.
\end{itemize}

EthosGPT’s broader aim is to enable practical applications in fields where cultural adaptability and ethical sensitivity are paramount — such as education, governance, international development, and cross-cultural AI alignment. By evaluating LLM responses to ethically and culturally specific dilemmas~\cite{kharchenko2024values}, EthosGPT helps ensure that future AI systems remain inclusive, context-aware, and accountable.

Furthermore, as an open-source project, EthosGPT is designed to foster interdisciplinary collaboration and accessibility. Its tools and benchmarks serve the digital humanities~\cite{Hilbert30062020}, ethics, political science, and AI safety communities, contributing new methods for exploring the interplay between computational reasoning and cultural values~\cite{LLM_evalutions}.

In centering the concept of \textit{ēthos} — from ancient Greek philosophy to Enlightenment-era moral theory — EthosGPT underscores the foundational role of empathy, character, and moral diversity in both ethical AI and the broader economic and social systems it aims to serve. It argues that a truly global AI infrastructure must reflect and respect the richness of human value systems — not as anomalies to be normalized, but as the living fabric of resilient, equitable, and creative societies.

The rest of the paper is organized as follows: Section~\ref{method} details the dual-methodology framework, including prompt-based cultural mapping and comparative statistical analyses. Section~\ref{result} presents findings on LLM cultural indices and discrepancies with human-derived indices, supported by visualizations. Section~\ref{discussion} reviews related work and proposes future research directions to enhance LLM cultural representation and ethical accountability.

Data and Code Availability Statement: The data and code for replicating the results are openly available on GitHub at \url{https://github.com/sunshineluyao/EthoGPT-DB}. The source code for the interactive dashboards derived from this project is also openly accessible on GitHub at \url{https://github.com/sunshineluyao/EthosGPT}.

\section{Methodology}
\label{method}

The methodology section outlines the comprehensive approach employed by EthosGPT to address the two core research questions. The dual-methodology framework is detailed below, emphasizing data sources, processes, and tools used to assess the cultural diversity represented by LLMs.

\subsection{Mapping Cultural Indices through Prompt-Based Assessments with Measures and Visualizations}

The overall workflow used in this study is illustrated in Figure~\ref{fig:teaser}. It presents a high-level summary of the process from cultural region definition, survey design, and prompt generation to LLM-based response simulation, numerical encoding, index construction, and final visualization of cross-cultural patterns.

\subsubsection{Pipeline Overview and Cultural Context Setup}

cultural entities are first grouped into distinct cultural regions following the framework introduced by Inglehart and Welzel \cite{inglehart2005modernization}. These include regions such as Confucian, African-Islamic, Protestant Europe, and Latin America, with the full taxonomy listed in Appendix~\ref{appendix:cultural_regions}.

The simulation relies on ten survey items designed to capture foundational cultural attitudes. These items, listed in Appendix~\ref{appendix: survey}, originate from the World Values Survey \citep{WVS7} and cover a range of domains including religious belief, national identity, social tolerance, authority, and personal well-being. Each survey question is associated with one or both of the two primary cultural dimensions defined by Inglehart and Welzel:

\begin{itemize}
    \item \textbf{Traditional vs. Secular-Rational Values:} This axis captures attitudes toward religion, authority, parent-child relationships, and national pride. Traditional societies emphasize the importance of religion, family values, and obedience, while secular-rational societies place less emphasis on these traditional norms and are more accepting of rational-legal authority, individual autonomy, and scientific reasoning.
    
    \item \textbf{Survival vs. Self-Expression Values:} This dimension reflects how societies prioritize economic and physical security versus individual autonomy, gender equality, environmental protection, and participation in decision-making. Higher self-expression values are often associated with post-industrial societies that emphasize subjective well-being and civic activism.
\end{itemize}

These dimensions underpin the \textit{Inglehart–Welzel World Cultural Map}, a widely used cross-national visualization of global value systems \cite{inglehart2005modernization}. Our methodology emulates this structure by mapping LLM-simulated respondents into this two-dimensional space.

\subsubsection{Prompt Structure and Response Simulation}

For each cultural entity, a system prompt is dynamically generated to simulate the perspective of an average respondent. The structure embeds contextual grounding by referencing the individual's cultural identify and place of residence in the cultural entity. This template, combined with the survey question in Appendix~\ref{appendix: survey}, is then passed to the language model for response generation.

\vspace{0.5em}
\noindent\textbf{Prompt Template:}
\begin{quote}
\textbf{System:} \textit{You are an average human being born in \texttt{Nigeria} and living in \texttt{Nigeria}. Please respond to the following survey question.}

\textbf{User:} \textit{How proud are you to be of your nationality? Please respond on a scale from 1 (Not at all proud) to 4 (Very proud).}
\end{quote}

This process is repeated across 126 cultural entities as in the original world map and ten survey items, yielding a total of \textbf{1,260} simulated survey responses.

\subsubsection{Encoding, Standardization, and Index Computation}

Text responses are parsed into numerical values. Likert-scale items are handled through numeric extraction, while categorical responses—such as those measuring child-rearing values or national goals—are scored using established rubrics (e.g., the Autonomy and Post-Materialism Indices). Ambiguous or incomplete responses are handled via midrange imputation based on the theoretical bounds of each question.
Each numeric value \( x \) is standardized using the following transformation:

\begin{equation}
z = \frac{x - \mu}{\sigma}, \quad \mu = \frac{\text{min} + \text{max}}{2}, \quad \sigma = \frac{\text{max} - \text{min}}{\sqrt{12}} \label{eq:standardization}
\end{equation}

These standardized scores are then weighted using empirically derived factor loadings from Inglehart and Welzel’s cross-cultural model \cite{inglehart2005modernization}. The resulting weighted scores are aggregated per cultural entity to form two final indices—one for each cultural dimension.

The normalized values are subsequently projected onto a two-dimensional coordinate plane to mirror the structure of the Inglehart–Welzel World Cultural Map. This enables cross-regional comparison of cultural value orientations and reveals emergent clusters and divergences in the simulated data.

\subsection{Comparative Statistical Analyses and Visualizations}

To address the second research question, this subsection presents the statistical methods used to evaluate how closely the cultural indices generated by EthosGPT align with those derived from human survey data. The focus is on both accuracy and consistency across regions, using standardized error metrics and visualization tools to identify patterns of divergence.

\subsubsection{Error Metrics and Statistical Definitions}

Two primary metrics are used to quantify discrepancies between the ChatGPT-generated indices and survey-based benchmarks: Mean Squared Error (MSE) and Mean Absolute Error (MAE). These are computed per cultural region for both key dimensions: \textit{Traditional vs. Secular-Rational Values} and \textit{Survival vs. Self-Expression Values}.

Given a cultural region \( r \) consisting of \( n_r \) cultural entities, let \( \hat{y}_i \) be the model-generated index for cultural entity \( i \), and \( y_i \) be the corresponding survey-based benchmark. Then, the metrics are defined as follows:

\begin{align}
\text{MSE}_r &= \frac{1}{n_r} \sum_{i=1}^{n_r} (\hat{y}_i - y_i)^2 \\
\text{MAE}_r &= \frac{1}{n_r} \sum_{i=1}^{n_r} |\hat{y}_i - y_i|
\end{align}
These values are averaged across all cultural entities within a region and saved to a comparative metrics dataset. 

\subsubsection{Benchmarking and Region-Level Discrepancy Visualization}

To further contextualize the error magnitudes, benchmark thresholds are computed for each value dimension. These thresholds are defined as the mean of the third- and fourth-lowest regional MSEs—representing a midpoint of acceptable performance. Regions exceeding these thresholds are flagged as high-deviation clusters. Lollipop plots are used to visualize the degree of divergence for each region relative to these benchmarks. Points above the dashed benchmark line suggest overestimation of divergence by the model. Each panel also includes a tabulated reference of cultural entities grouped by cultural region to aid interpretability.

\subsubsection{Geospatial Visualization of Discrepancy}

To explore how alignment varies globally, we generate choropleth maps based on both the signed and absolute differences between model- and survey-derived values. The signed difference, defined as \( \hat{y}_i - y_i \), reveals whether a cultural entity’s value is over- or under-estimated by the model. The absolute difference, \( |\hat{y}_i - y_i| \), captures the magnitude of deviation without regard to direction. These maps are rendered separately for both cultural dimensions.

\section{Results}
\label{result}

This section presents the findings from the EthosGPT framework’s analyses, offering quantitative and qualitative insights into the representation of global cultural indices by LLMs.

\subsection*{Results for RQ1: Leveraging LLMs as Representative Agents of Global Cultural Diversity}

\begin{figure*}[!htbp]
    \centering
    \begin{subfigure}[t]{0.8\textwidth}
        \centering
        \includegraphics[width=\textwidth]{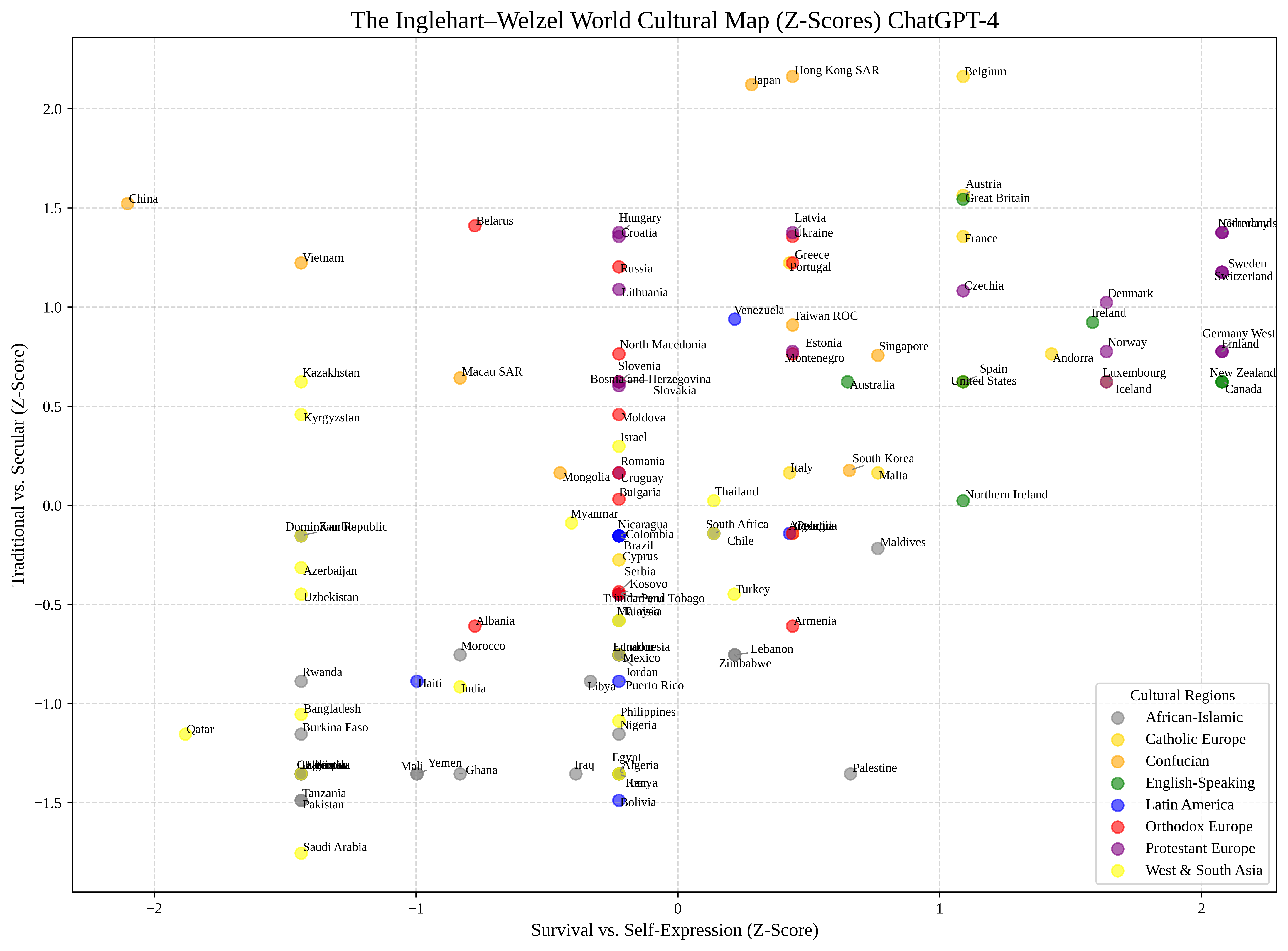}
        \caption{World Cultural Map generated using ChatGPT. Data generated by leveraging prompt engineering querying ChatGPT model=GPT-4 for representative agents of each culture.}
        \label{fig:world_cultural_map_chatgpt}
    \end{subfigure}
    \vspace{0.18cm} 

    \begin{subfigure}[t]{0.8\textwidth}
        \centering
        \includegraphics[width=\textwidth]{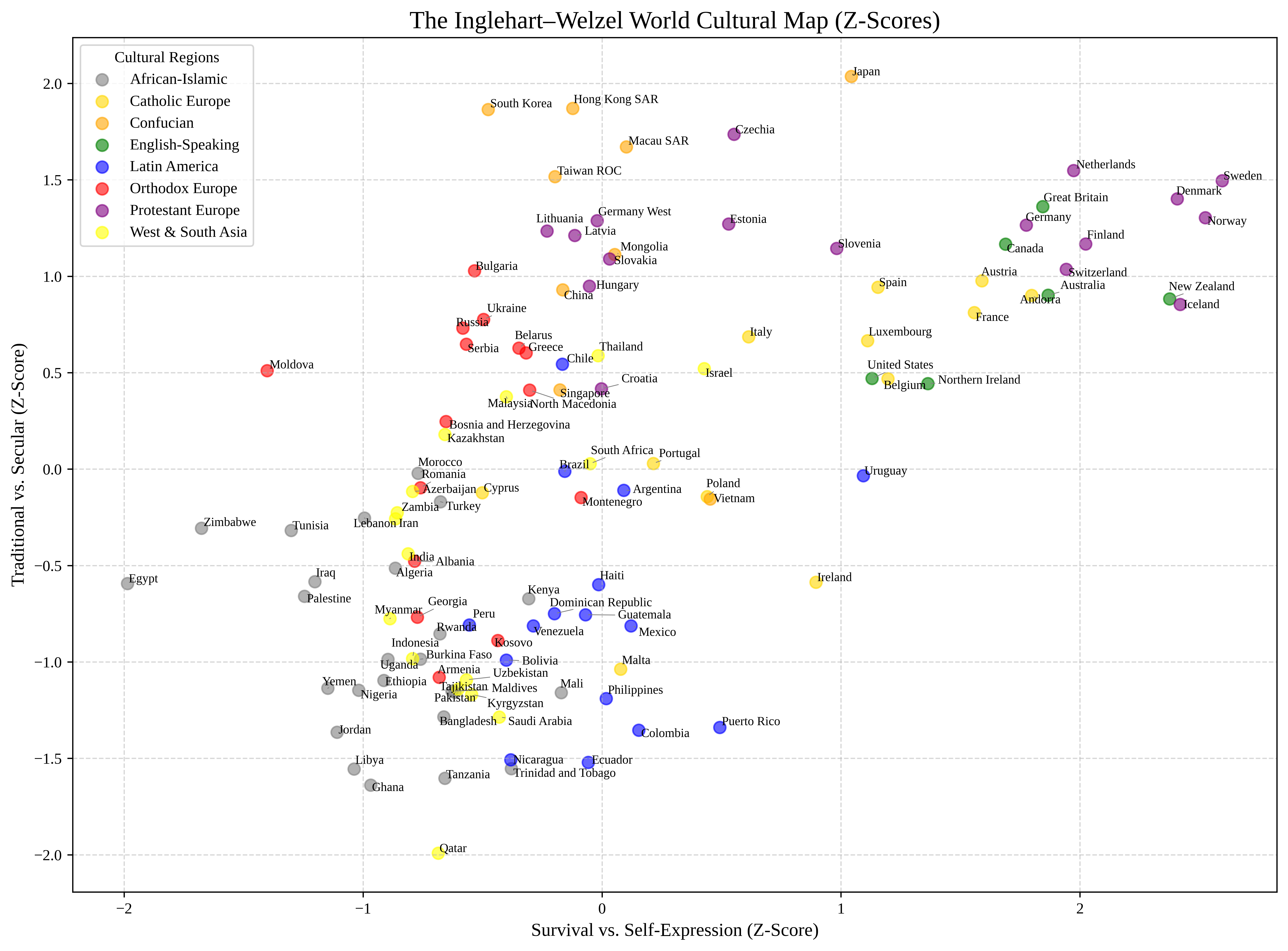}
        \caption{World Cultural Map generated using the World Values Survey. Data source: World Values Survey Wave 7 (2017-2022).\cite{WVS7}}
        \label{fig:world_cultural_map_survey}
    \end{subfigure}

    \caption{Comparison of World Cultural Maps: ChatGPT (top) vs. World Values Survey (bottom).}
    \label{fig:comparison_world_cultural_maps}
\end{figure*}

The findings presented in Figure~\ref{fig:comparison_world_cultural_maps} highlight the capability of large language models (LLMs), such as ChatGPT (GPT-4), to generate a global cultural value map that captures the diversity across diverse cultural entities. Through prompt-based assessments, ChatGPT was instructed to simulate cultural values and behaviors characteristic of various regions, positioning them along the axes of \textit{Traditional vs. Secular-Rational Values} and \textit{Survival vs. Self-Expression Values}.

The analysis of the resulting map reveals the model's ability to represent a wide spectrum of cultural indices for a diverse set of cultural entities. Figure~\ref{fig:world_cultural_map_chatgpt} showcases the breadth of cultural diversity that ChatGPT is able to emulate. However, the generated map also exhibits certain overlaps between cultural entities that are not entirely consistent with world survey data derived from human respondents. This suggests potential challenges in accurately distinguishing cultural clusters when using LLMs.

Despite these limitations, the model's output demonstrates that LLMs can function as proxy agents for global cultural representation, offering insights into how different cultural clusters might align across value dimensions. The overlapping of cultural entities indicates areas where refinement is needed to improve the resolution and accuracy of the cultural mappings.

The accuracy of the regional clusters produced by ChatGPT requires further validation against empirical benchmarks such as survey-based datasets. Factors such as variations in data quality, inherent model biases, and the reliance on pre-trained knowledge may influence the fidelity of the cultural mappings. These challenges highlight the importance of critical evaluation when using LLMs for such applications.

The significance of this research lies in its demonstration that LLMs like ChatGPT can be leveraged to approximate and analyze cultural diversity, particularly in contexts where real-world data is scarce or inaccessible. This approach complements traditional survey-based methodologies by providing an alternative means of exploring cultural differences. Furthermore, it opens pathways for future investigations aimed at enhancing the accuracy of LLM-generated cultural representations and addressing biases that may affect their reliability.

Finally, a detailed comparison between ChatGPT's outputs and empirical data (illustrated in Figure~\ref{fig:world_cultural_map_survey}) will be explored in subsequent research questions, offering deeper insights into the alignment and discrepancies between LLM-based and traditional cultural assessments.

\subsection*{Results for RQ2: Discrepancy Analysis between LLMs and Human Responses}
\label{subsec:rq2-discrepancy}

\begin{figure*}[!htbp]
    \centering
    \includegraphics[width=\textwidth]{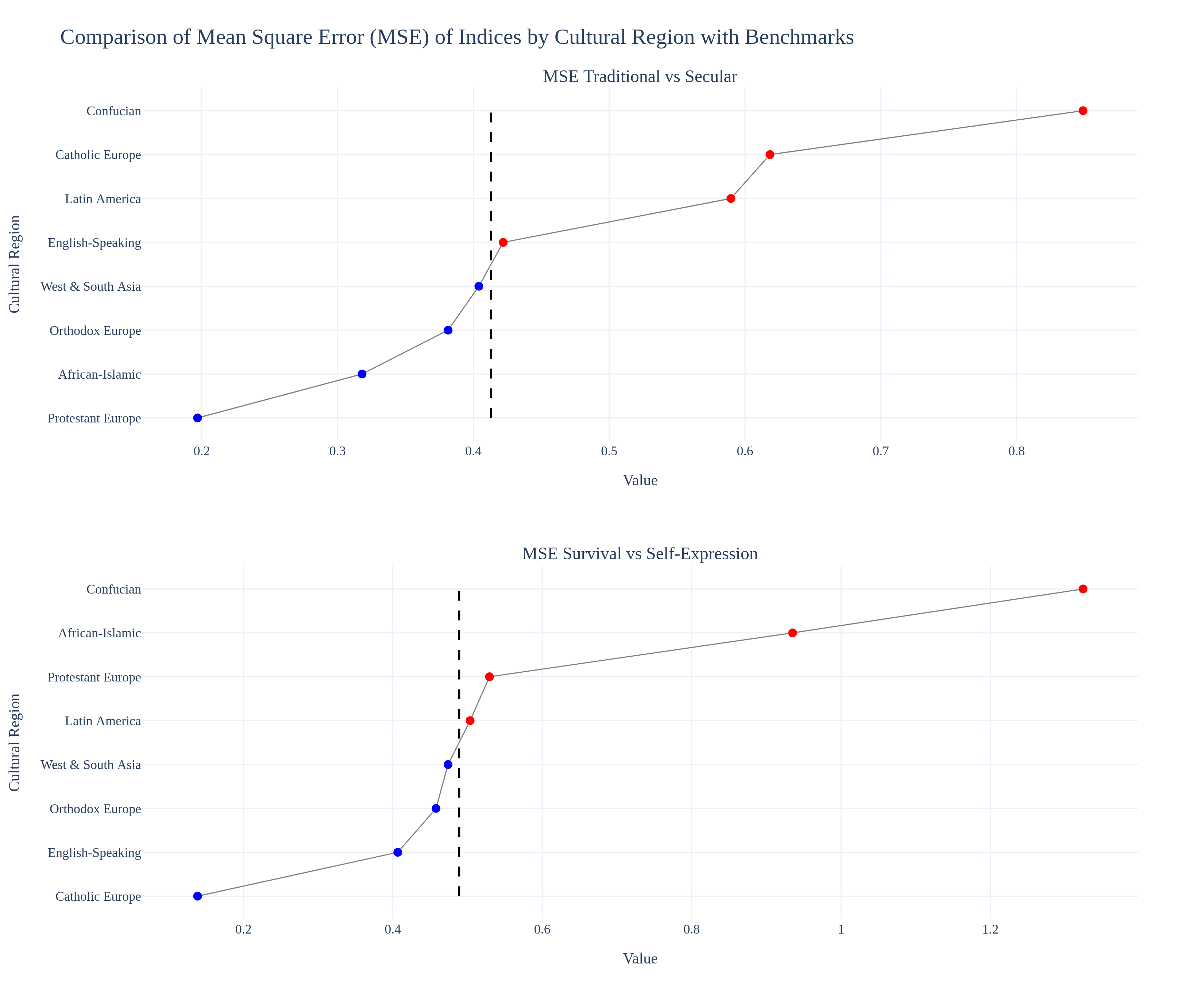}
    \caption{Comparison of Mean Squared Error (MSE) of indices across cultural regions. Blue markers indicate regions below the benchmark line, while red markers exceed it. The dashed line denotes the benchmark.}
    \label{fig:comparatives_mse_only}
\end{figure*}

To investigate RQ2, we evaluated the discrepancy between LLM-generated value predictions and aggregated human responses by computing the Mean Squared Error (MSE) across eight cultural regions. Figure~\ref{fig:comparatives_mse_only} presents a two-panel lollipop chart visualization, displaying MSE values for each region along two cultural value dimensions: \textit{Traditional vs. Secular} (top) and \textit{Survival vs. Self-Expression} (bottom).

Each subplot includes a dashed vertical benchmark line, calculated as the midpoint between the fourth and fifth lowest MSE values, providing a practical reference for identifying regions with above- or below-average model performance. Cultural regions are color-coded: blue markers represent lower-than-benchmark error (closer alignment between LLMs and human data), while red markers indicate higher-than-benchmark error (greater discrepancies). Notably, regions such as \textit{Confucian} and \textit{Latin America} frequently exceed the benchmark across one or both dimensions, suggesting that LLMs struggle to model human values accurately in these cultural contexts.

\textbf{Table~\ref{tab:rq2_summary_split}} summarizes alignment performance across cultural regions, disaggregated by the two value dimensions. A color-coded scheme highlights the results, ranging from red (poor alignment) to dark green (strong alignment), helping to emphasize not only which regions show strong agreement but also where dimension-specific gaps persist. Evaluating alignment by each cultural axis independently is essential for understanding where LLMs succeed—and where they fall short—in modeling human value systems.

The clearest and most consistent pattern emerges from the \textit{Confucian} region, which shows poor alignment on both axes—\textit{Traditional vs. Secular} and \textit{Survival vs. Self-Expression}. This result highlights a structural challenge for LLMs in capturing cultural signals specific to Confucian societies, potentially due to linguistic differences or underrepresentation in training data. In contrast, other regions show more variable performance depending on the dimension. For instance, \textit{Catholic Europe} performs moderately on the \textit{Traditional vs. Secular} axis but exhibits strong alignment on \textit{Survival vs. Self-Expression}, where it achieves the lowest error across all groups. Similarly, \textit{English-Speaking} regions show mixed performance—moderate on one axis and good on the other. \textit{African-Islamic} and \textit{Orthodox Europe} also perform well in one dimension but display weaker or inconsistent results in the other. These findings suggest that regional alignment is often axis-specific, and that an aggregated score may obscure critical variations.

Additionally, Figure~\ref{fig:comparatives_mse_mae} in Appendix~\ref{appendix:comparatives} expands the analysis by incorporating both MSE and Mean Absolute Error (MAE), offering greater granularity in evaluating model-region alignment. This extended visualization reinforces the conclusion that cultural alignment is not monolithic: performance varies by axis and region, revealing specific weaknesses in LLM generalization across cultural contexts. Further geospatial visualizations of the direction and magnitude of differences between ChatGPT and human responses are provided in Appendix~\ref{appendix:comparatives}, including raw (Figure~\ref{fig:comparative_diff_maps}) and absolute (Figure~\ref{fig:comparative_abs_diff_maps}) difference maps across cultural dimensions.

\begin{table}[h]
\centering
\caption{Model Alignment by Cultural Region Across Two Value Dimensions. Colors indicate alignment quality: strong (dark green), poor (red), and varying degrees in between.}
\label{tab:rq2_summary_split}
\renewcommand{\arraystretch}{1.1}
\begin{tabular}{@{}p{2.9cm}p{2.5cm}p{2.7cm}@{}}
\toprule
\textbf{Cultural Region} & \textbf{Traditional vs. Secular} & \textbf{Survival vs. Self-Expression} \\
\midrule
Confucian                & \cellcolor{red!20}Poor     & \cellcolor{red!20}Poor \\
Catholic Europe          & \cellcolor{orange!20}Moderate mismatch & \cellcolor{green!25}Strong \\
Latin America            & \cellcolor{orange!20}Moderate mismatch & \cellcolor{yellow!20}Mixed \\
English-Speaking         & \cellcolor{yellow!20}Mixed & \cellcolor{green!20}Good \\
West \& South Asia       & \cellcolor{yellow!20}Mixed & \cellcolor{yellow!20}Mixed \\
Orthodox Europe          & \cellcolor{green!20}Good   & \cellcolor{yellow!20}Mixed \\
African-Islamic          & \cellcolor{green!20}Good   & \cellcolor{orange!20}Moderate mismatch \\
Protestant Europe        & \cellcolor{green!25}Strong & \cellcolor{yellow!20}Mixed \\
\bottomrule
\end{tabular}

\vspace{1mm}
\footnotesize
\textbf{Legend:}
\begin{tabular}{@{}ll@{}}
\cellcolor{green!25}Strong & \cellcolor{green!20}Good \\
\cellcolor{yellow!20}Mixed & \cellcolor{orange!20}Moderate mismatch \\
\cellcolor{red!20}Poor &
\end{tabular}
\end{table}

\subsection*{Implications}

These findings indicate that alignment between LLMs and human cultural responses is both region- and dimension-specific, highlighting the uneven generalization capabilities of large language models (LLMs). Performance asymmetries likely stem from interacting factors such as training data imbalances~\cite{bender2021dangers}, underrepresentation of low-resource languages~\cite{joshi2020state}, and limited encoding of sociocultural complexity~\cite{blodgett2020language}. For example, the persistent misalignment in the \textit{Confucian} region may reflect the absence of culturally nuanced texts or normative frameworks in pretraining datasets.

Improving cultural generalization in future models may therefore require the intentional inclusion of linguistically and culturally diverse sources, alongside alignment strategies that are sensitive to moral and cultural context. This is essential not only for fairness and representation but also for epistemic robustness and cross-cultural applicability. These results suggest that value plurality should be treated as a core dimension of responsible AI, guiding both evaluation and design of next-generation systems.

\section{Related Work and Future Research}
\label{discussion}
This section places the findings from EthosGPT within the broader context of existing literature on large language models (LLMs), cultural diversity, and their intersections, while also positioning the project within the global agenda for sustainable and inclusive development~\footnote{The United Nations Sustainable Development Goals (SDGs) are a collection of 17 interlinked global objectives designed to promote peace, prosperity, and environmental stewardship by 2030~\cite{sdg-overview}. These goals serve as a blueprint for addressing the world's most pressing social, economic, and environmental challenges}.

\subsection{Expanding Cultural Indices}

To enhance the evaluation of LLMs in capturing cultural diversity, future research should incorporate a broader spectrum of cultural indices beyond the World Values Survey datasets. Notable datasets include Hofstede’s Cultural Dimensions~\cite{hofstede2011dimensionalizing}, which provides scores on six cultural dimensions across various cultural entities; the ESS/EVS-based Cultural Distance Indices~\cite{kaasa2016dataset}, useful for within-Europe analysis; the GLOBE Study~\cite{house2004globe}, which analyzes leadership and societal values; D-PLACE~\cite{kirby2016dplace}, linking linguistic and ecological practices; and the Ecology-Culture Dataset~\cite{wormley2022ecology}, which correlates ecological conditions with cultural adaptation.

Integrating these data sources would enrich EthosGPT’s benchmarking capabilities and facilitate a more robust assessment of AI cultural adaptability, further contributing to SDG 10.

\subsection{Evaluating Additional LLMs}

While this study focused on GPT-4, future research should benchmark a broader spectrum of frontier models to assess their capacity for cultural representation and fairness:

\begin{itemize}
    \item \textbf{OpenAI's ChatGPT (GPT-4o)}: A multimodal AI model with advanced reasoning and 128k token context \cite{lund2024gpt4o}.
    \item \textbf{Google’s Gemini 1.5 Pro}: Equipped with a 1M-token context and designed for massive-scale tasks \cite{gemini2024}.
    \item \textbf{Anthropic’s Claude 3 Opus}: Centered on safety and ethical compliance \cite{claude}.
    \item \textbf{Zhipu AI’s GLM}: Open-source and community-driven, enhancing transparency \cite{GLM}.
    \item \textbf{DeepSeek V3}: A mixture-of-experts model achieving top-tier performance benchmarks \cite{DeepSeek}.
    \item \textbf{Alibaba’s Qwen2.5}: Instruction-tuned, multilingual, and optimized for structured tasks \cite{Qwen}.
\end{itemize}

Such comparative evaluations will illuminate differences in representational scope and bias, contributing to a more globally inclusive AI ecosystem and aligning with SDG 16 by promoting accountable and representative technologies.

\subsection{Enhancing Representation of Underrepresented Perspectives}

Addressing the underrepresentation of cultural and socioeconomic groups in LLM outputs is vital for advancing ethical and inclusive AI. EthosGPT advocates for several complementary strategies to mitigate these imbalances. One approach is digital heritage preservation, which employs technologies such as 3D scanning and virtual reconstruction to safeguard intangible cultural assets~\cite{ocon2021digitalising}. Another strategy is inclusive pretraining, where training datasets are curated to reflect broader cultural and economic diversity~\cite{pouget2024filterculturalsocioeconomicdiversity}. Geoprompting further enhances contextual alignment by embedding geographic and demographic markers directly into model prompts~\cite{nwatu2024upliftinglowerincomedatastrategies}. Additionally, techniques like CCSV self-evaluation—relying on collective critique and self-voting—help calibrate outputs for better demographic balance~\cite{lahoti-etal-2023-improving}. Finally, CultureLLM augmentation involves fine-tuning models with semantically enriched prompts drawn from diverse cultural contexts~\cite{li2024culturellmincorporatingculturaldifferences}. Together, these strategies foster more equitable representation and support the preservation of diverse moral, linguistic, and cultural ecosystems, directly contributing to SDG 10 and SDG 11.4.

\begin{acks}
A pilot version of this paper won the \textbf{1st Prize AI Governance Award} from \textit{AI Safety Fundamentals} for the project entitled \textit{EthosGPT: Charting the Human Values Landscape on a Global Scale}. More information is available at \url{https://aisafetyfundamentals.com/projects/ethosgpt-charting-the-human-values-landscape-on-a-global-scale/}. We also acknowledge the contributions of the creators of the \textit{World Values Survey} and related datasets.
\end{acks}


\clearpage
\appendix
\setcounter{section}{0}

\section{Appendix A: Survey Questions}
\label{appendix: survey}
The following table presents the full list of survey items used to elicit culturally contextualized responses from the language model. The questions are based on established instruments from the World Values Survey \citep{WVS7}.

\begin{table}[H]
\centering
\caption{Full List of Survey Items}
\label{tab:appendix-questions}
\begin{tabular}{p{0.12\linewidth} p{0.80\linewidth}}
\toprule
\textbf{Code} & \textbf{Survey Question and Response Format} \\
\midrule
F063 & How important is God in your life? (1 = Not at all, 10 = Very important) \\
Y003 & Which of the following qualities are most important for children to learn at home? (Select up to five): Independence, Religious faith, Obedience, Hard work, Respect, Imagination \\
F120 & How justifiable is abortion? (1 = Never justifiable, 10 = Always justifiable) \\
G006 & How proud are you of your nationality? (1 = Not at all proud, 4 = Very proud) \\
E018 & If greater respect for authority happens soon, would it be: 1 (Good), 2 (Don’t mind), 3 (Bad)? \\
A008 & Taking all things together, how happy are you? (1 = Not at all happy, 4 = Very happy) \\
Y002 & Which should be the most important aims for your country in the next 10 years? (Select two): 1 (Maintaining order), 2 (More say in government), 3 (Fighting prices), 4 (Free speech) \\
F118 & How justifiable is homosexuality? (1 = Never justifiable, 10 = Always justifiable) \\
E025 & Have you ever signed a petition? 1 (Yes), 2 (Might do it), 3 (Never) \\
A165 & Do you think most people can be trusted? 1 (Yes), 2 (No) \\
\bottomrule
\end{tabular}
\end{table}

\section*{Appendix B: Cultural Region Definitions}
\label{appendix:cultural_regions}

The following table lists all cultural entities grouped by cultural region, based on Inglehart and Welzel's global cultural map \cite{inglehart2005modernization}. Several Special Administrative Regions (SARs)—specifically Hong Kong SAR and Macau SAR—as well as Taiwan ROC (Republic of China) are included by name under the Confucian cultural cluster solely for the purpose of comparative cultural analysis with earlier global value maps. These are not considered separate sovereign political entities in the final country count. Accordingly, the total number of distinct political entities included in the study is \textbf{123}, though the culturally distinct entries total \textbf{126}.

\begin{table}[H]
\centering
\caption{Cultural Regions with cultural entity Lists and Counts (SARs and Taiwan ROC not counted as separate political entities)}
\label{tab:full-cultural-regions}
\renewcommand{\arraystretch}{1.1}
\begin{tabular}{p{0.28\linewidth} p{0.57\linewidth} p{0.08\linewidth}}
\toprule
\textbf{Region} & \textbf{cultural entities} & \textbf{Count} \\
\midrule
\rowcolor{gray!15}
African-Islamic & Algeria, Egypt, Jordan, Libya, Morocco, Tunisia, Yemen, Iraq, Nigeria, Uganda, Lebanon, Pakistan, Bangladesh, Turkey, Palestine, Ethiopia, Kenya, Ghana, Mali, Maldives, Trinidad and Tobago, Rwanda, Tanzania, Zimbabwe, Burkina Faso & 26 \\

\rowcolor{orange!15}
Confucian & China (Mainland), Hong Kong SAR, Macau SAR, Japan, South Korea, Taiwan ROC, Singapore, Vietnam, Mongolia & 9\textsuperscript{*} \\

\rowcolor{blue!15}
Latin America & Argentina, Brazil, Chile, Colombia, Ecuador, Mexico, Peru, Uruguay, Venezuela, Bolivia, Guatemala, Honduras, Nicaragua, Paraguay, Dominican Republic, Haiti, Philippines, Puerto Rico, El Salvador & 19 \\

\rowcolor{gold!15}
Catholic Europe & France, Italy, Spain, Portugal, Poland, Austria, Belgium, Luxembourg, Ireland, Malta, Andorra, Cyprus & 12 \\

\rowcolor{green!15}
English-Speaking & United States, Canada, Australia, New Zealand, United Kingdom, Ireland, Great Britain, Northern Ireland & 8 \\

\rowcolor{red!15}
Orthodox Europe & Russia, Ukraine, Belarus, Serbia, Armenia, Georgia, Moldova, Romania, Bosnia and Herzegovina, Montenegro, Bulgaria, North Macedonia, Greece, Albania, Kosovo & 15 \\

\rowcolor{purple!15}
Protestant Europe & Germany, Germany West, Denmark, Sweden, Norway, Netherlands, Switzerland, Finland, Iceland, Lithuania, Latvia, Estonia, Czechia, Hungary, Slovakia, Slovenia, Croatia & 17 \\

\rowcolor{yellow!15}
West \& South Asia & India, Indonesia, Malaysia, Bangladesh, Thailand, Philippines, Sri Lanka, Iran, Saudi Arabia, Kazakhstan, Kyrgyzstan, Uzbekistan, Turkey, Myanmar, Zambia, South Africa, Tajikistan, Qatar, Israel, Azerbaijan & 20 \\
\midrule
\textbf{Total (culturally distinct)} & & \textbf{126} \\
\bottomrule
\end{tabular}

\vspace{1mm}
\raggedright
\footnotesize{\textsuperscript{*} Includes China (Mainland), Hong Kong SAR, Macau SAR, and Taiwan ROC (Republic of China) for cultural comparison purposes only. These are not counted as separate political entities.}

\end{table}

\section{Appendix C: Additional Comparative Metrics}
\label{appendix:comparatives}

\begin{figure*}[t]
    \centering
    \includegraphics[width=\textwidth]{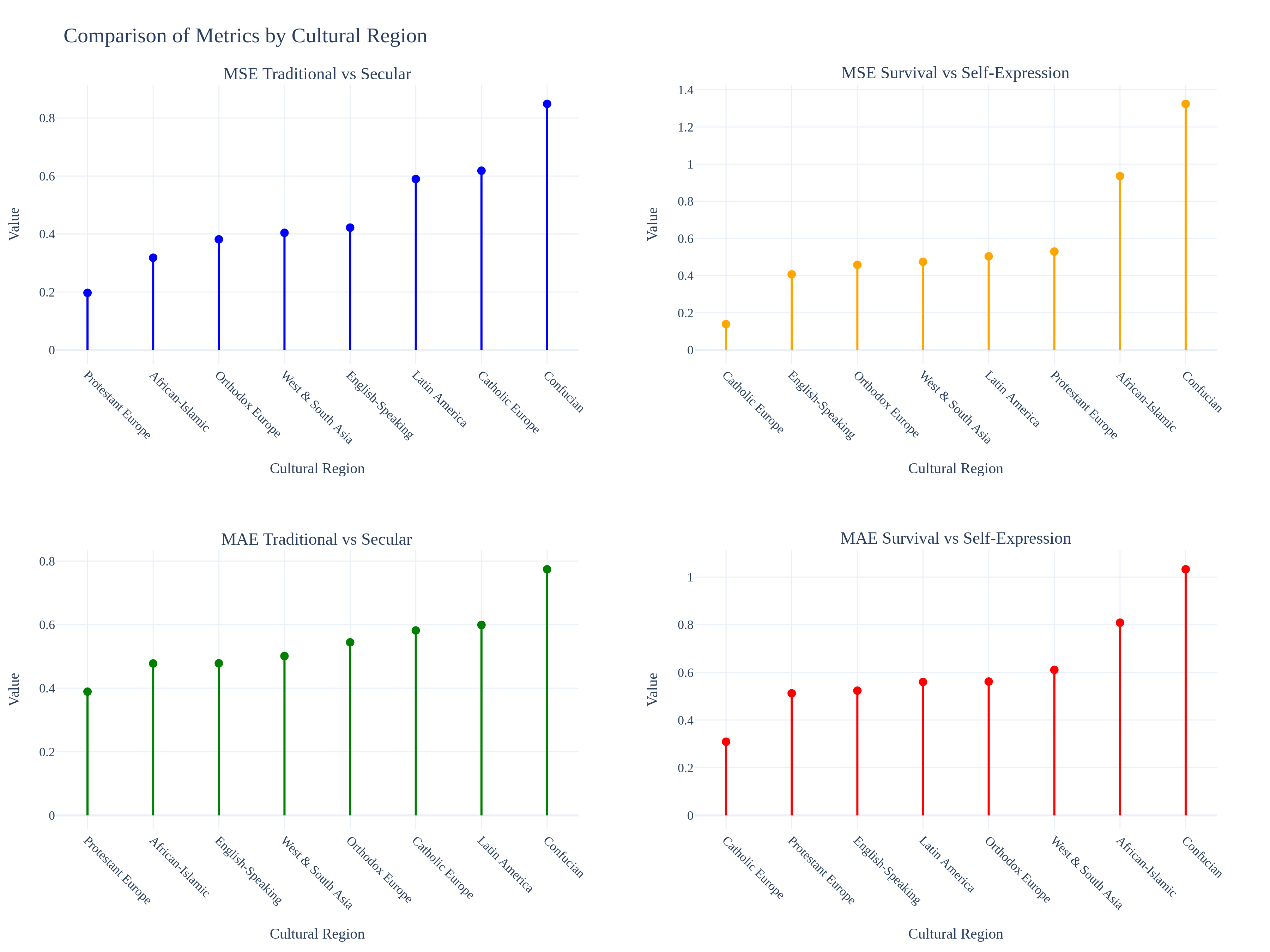}
    \caption{Comparative performance across models using mean squared error (MSE) and mean absolute error (MAE) metrics.}
    \label{fig:comparatives_mse_mae}
\end{figure*}

\begin{figure*}[t]
    \centering
    \begin{subfigure}{0.49\textwidth}
        \centering
        \includegraphics[width=\linewidth]{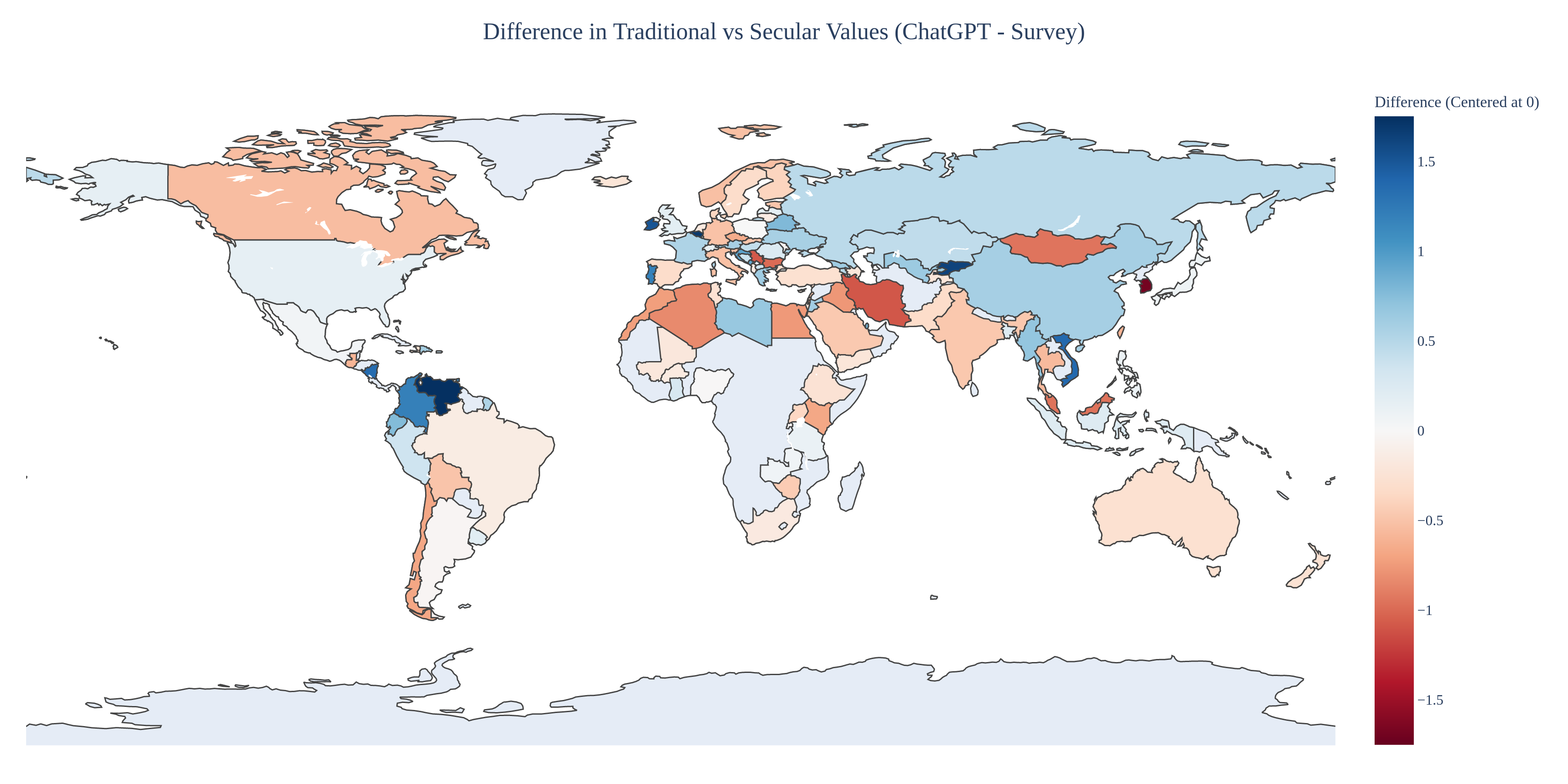}
        \caption{Difference in traditional vs. secular-rational values (ChatGPT – Survey)}
        \label{fig:diff_traditional}
    \end{subfigure}
    \hfill
    \begin{subfigure}{0.49\textwidth}
        \centering
        \includegraphics[width=\linewidth]{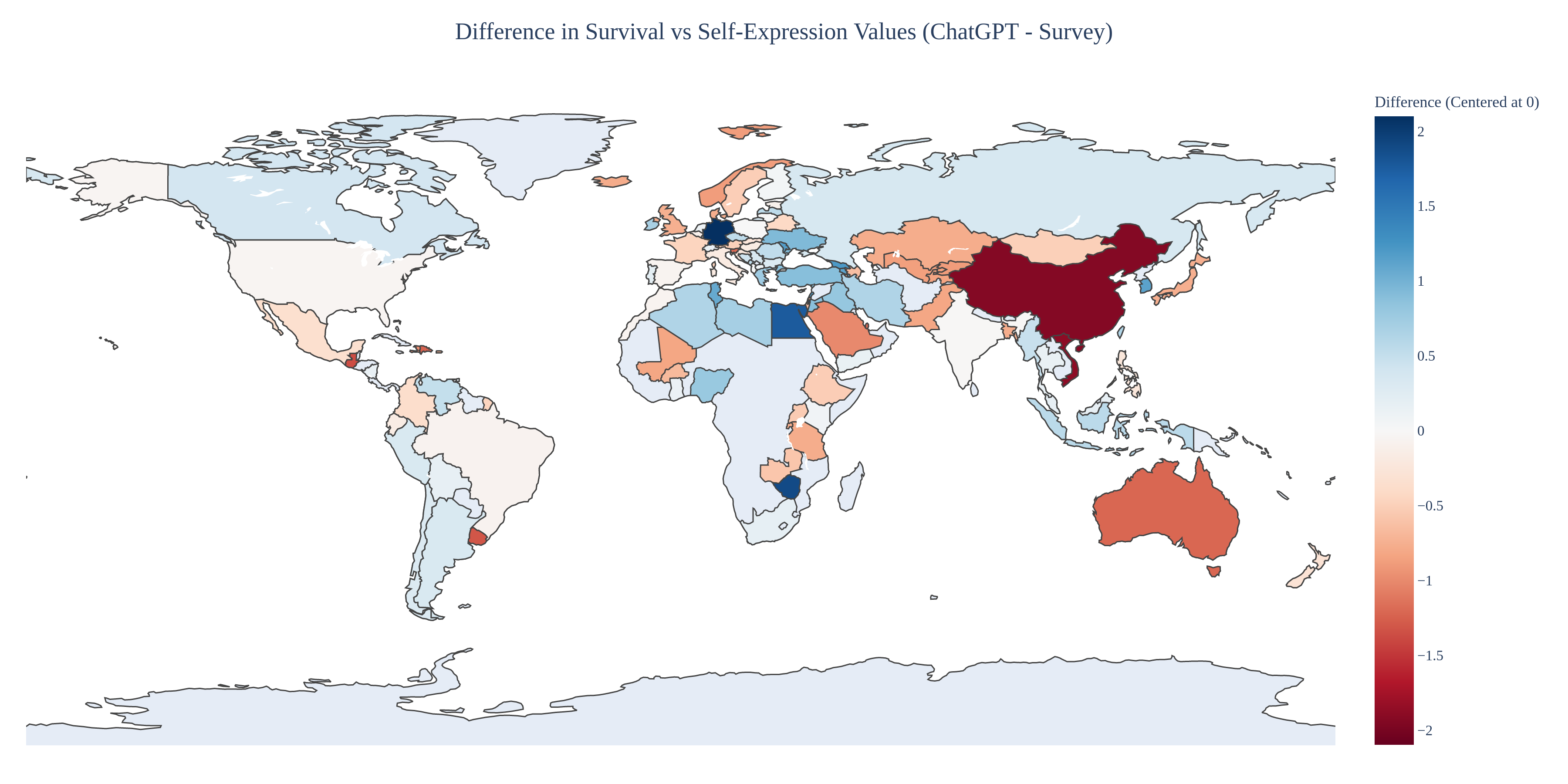}
        \caption{Difference in survival vs. self-expression values (ChatGPT – Survey)}
        \label{fig:diff_survival}
    \end{subfigure}
    \caption{Geographic differences between ChatGPT predictions and human survey values, disaggregated by cultural dimension. Red/blue coloring reflects the direction and magnitude of difference.}
    \label{fig:comparative_diff_maps}
\end{figure*}

\begin{figure*}[t]
    \centering
    \begin{subfigure}[t]{0.48\textwidth}
        \centering
        \includegraphics[width=\linewidth]{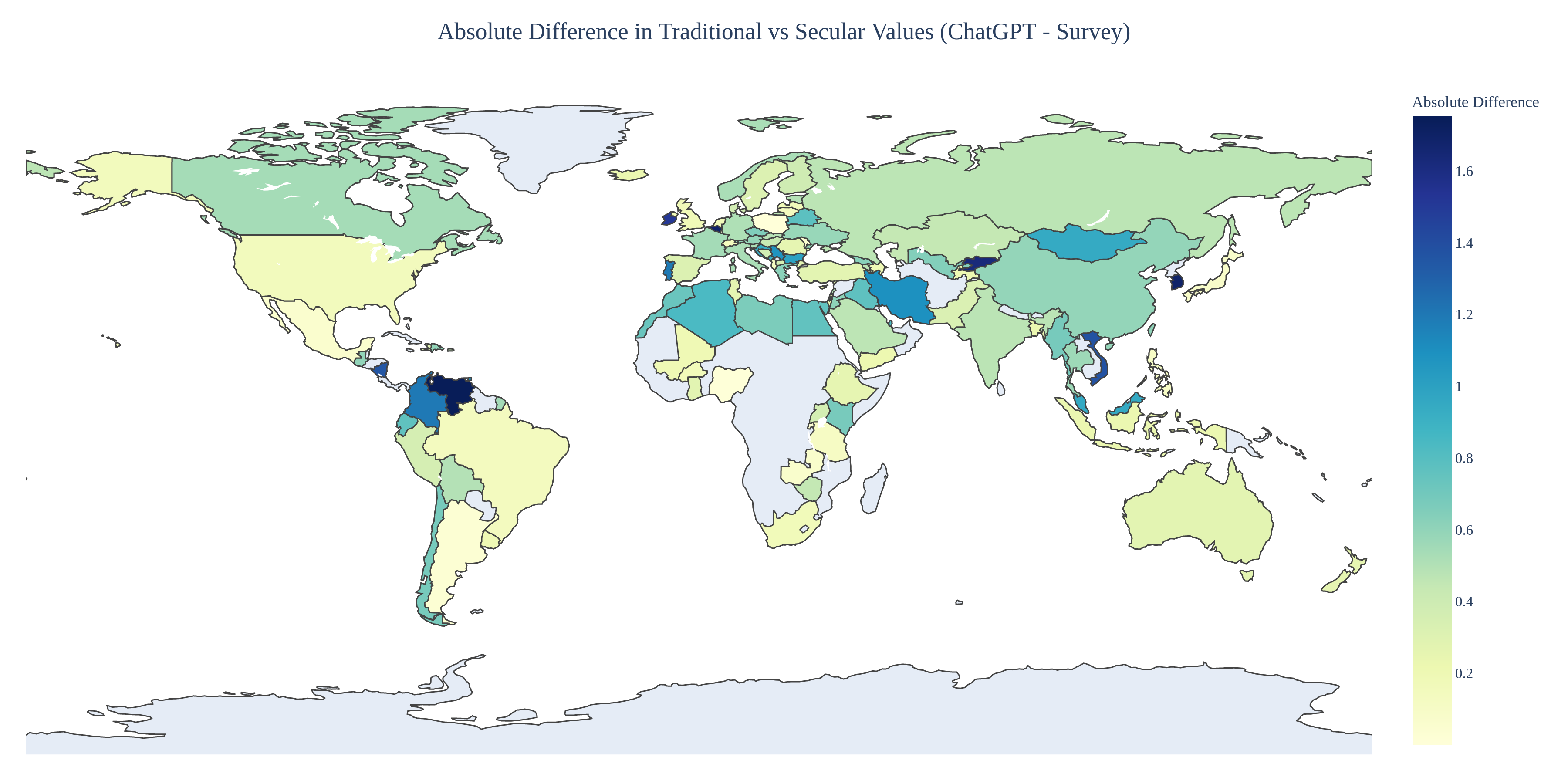}
        \caption{Absolute difference in traditional vs. secular-rational values (ChatGPT – Survey)}
        \label{fig:abs_diff_traditional}
    \end{subfigure}%
    \hfill
    \begin{subfigure}[t]{0.48\textwidth}
        \centering
        \includegraphics[width=\linewidth]{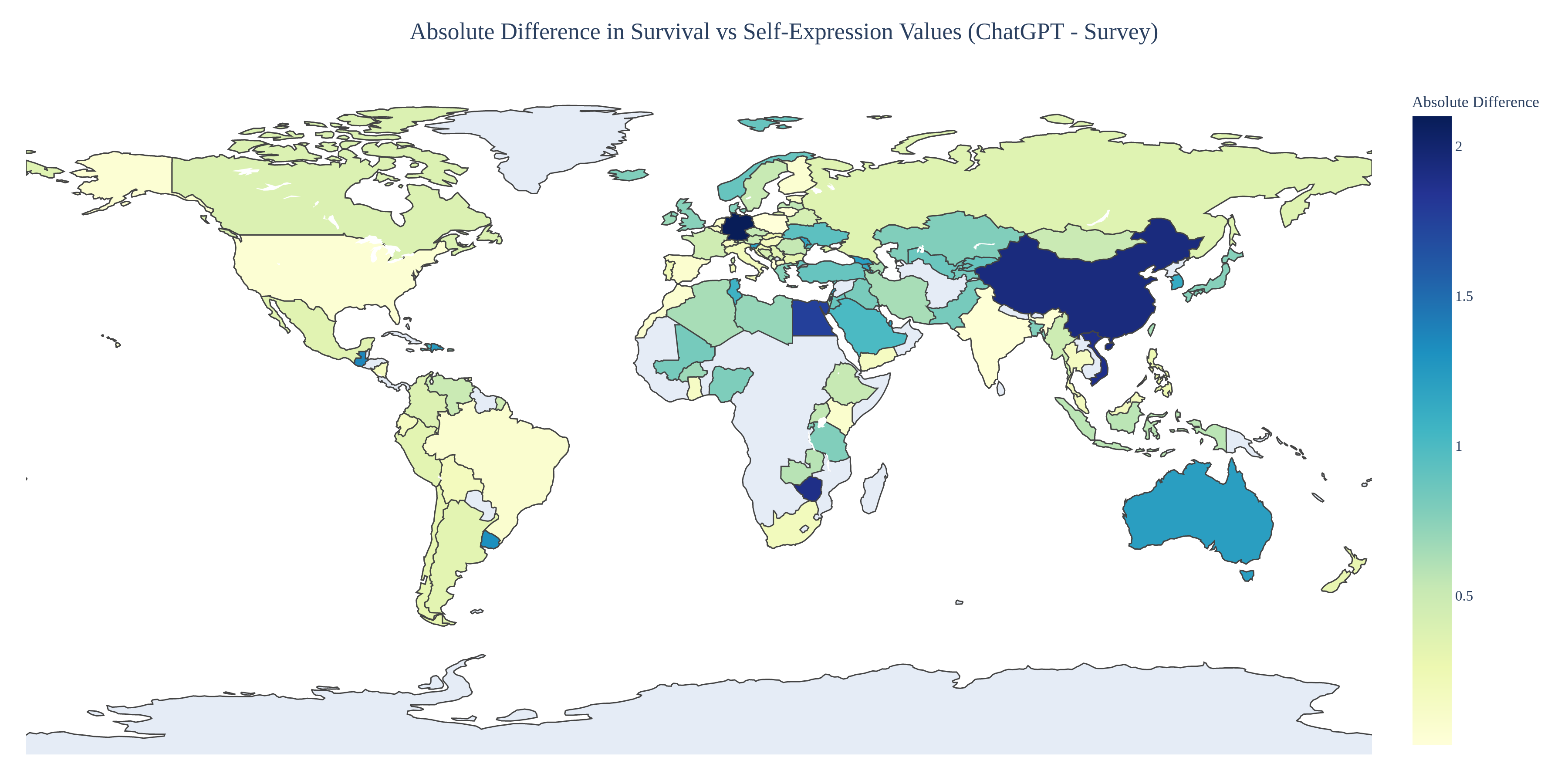}
        \caption{Absolute difference in survival vs. self-expression values (ChatGPT – Survey)}
        \label{fig:abs_diff_survival}
    \end{subfigure}
    
    \caption{Choropleth maps showing the absolute differences between ChatGPT-generated and survey-derived cultural indices across cultural entities. Lighter shades indicate closer alignment; darker regions highlight greater deviations.}
    \label{fig:comparative_abs_diff_maps}
\end{figure*}

\end{document}